\documentclass{article}
\usepackage[utf8]{inputenc}
\usepackage[table]{xcolor}
\usepackage{amsmath}
\usepackage{amstext}
\usepackage{csquotes}
\usepackage{collcell}
\usepackage{hhline}
\usepackage{hyperref}
\usepackage{pgf}
\usepackage{multicol}
\usepackage[margin=1.0in]{geometry}
\pdfoutput=1
\title{\textbf{On the Challenges of Detecting Rude Conversational Behaviour}}

\begin{document}

\author{
  Karan Grewal, Khai N. Truong\\
  Department of Computer Science\\
  University of Toronto\\
  \texttt{karanraj.grewal@mail.utoronto.ca}, \texttt{khai@cs.toronto.edu}
}
\date{}

\maketitle

\begin{abstract}
In this study, we aim to identify moments of rudeness between two individuals.
In particular, we segment all occurrences of rudeness in conversations into three broad, distinct categories and try to identify each.
We show how machine learning algorithms can be used to identify rudeness based on acoustic and semantic signals extracted from conversations.
Furthermore, we make note of our shortcomings in this task and highlight what makes this problem inherently difficult.
Finally, we provide next steps which are needed to ensure further success in identifying rudeness in conversations.
\end{abstract}

\section{Introduction}
One-on-one interactions are important in everyday social settings.
For instance, in order to attract a potential partner, it is imperative that an individual behave in an appropriate manner.
Unfortunately, one-on-one interactions can often result in one party exhibiting rude or inappropriate conversational behaviour.
In many cases, the offending party is not aware of the severity of their actions and does not intend to offend the other party.
For example, certain individuals may be socially unaware of how others perceive their behaviour.
Individuals with learning disabilities, such as autism, may follow this trend.
Likewise, young children often lack awareness of their behaviour -- a possible explanation for the presence of bullying in elementary schools and why children are generally regarded as immature.
In both cases, monitoring a user's conversational behaviour and making them aware of it via active feedback while they are engaged in a one-on-one interaction would be helpful towards correcting their behaviour in such scenarios.

In the last century, there has been a lot of work in the linguistics and psychology domains which attempt to define politeness and acceptable behaviour pertaining to two-person interactions.
The most popular of these is Penelope Brown and Steven Levinson's Politeness theory~\cite{brownlevinson}.
This theory states that all individuals have two \textit{faces}: a positive self-image which is the desire to be approved by others, and a negative self-image which is the desire of actions to be unimpeded by others.
According to Politeness theory, any external actions which threaten one or more of an individual's faces, such disrespectful gestures, constitute impoliteness.
Also, Geoffrey Leech's principle of politeness states that if two individuals are interacting, then there will be some form of disagreement or tension if both individuals are pursuing mutually-incompatible goals -- likening the chance of rude behaviour~\cite{leech}.
Here, \textit{goals} refers to a psychological state of being.
In contrast, Bruce Fraser argues against the theories formulated by Leech, Brown, and Levinson by pointing out that each culture has its own set of social norms which define acceptable behaviour~\cite{fraser}.
Therefore, as Fraser argues, the question of whether an individual is behaving in an inappropriate manner is entirely dependent on the context of his/her actions.
This view aligns with Robin Lakoff's notable example of the speaking style in New York~\cite{lakoff}.
As she states, New Yorkers often use profanity in a casual sense without any intent to offend or be impolite.
However, their conversational behaviour is likely to be interpreted as rude in other cultures.

Is there a grounded definition of rudeness with respect to speech which can be derived from classical theories of politeness?
In this study, we define define the notion of \textit{rude conversational behaviour} and explore methods to identify this type of behaviour in two-person interactions.
We do this by extracting acoustic and semantic information from an individual's speech and develop methods which attempt to pinpoint exact instances of rude conversational behaviour.
Also, we highlight some existing problems which make the task at hand difficult through our findings.
Note that we only focus on signals extracted speech data.

\section{Related Work}
The broader goal of identifying rude conversational behaviour is composed of subtasks which contribute to the larger goal by determining if some criteria for rude conversational behaviour is met.
Of these, sentiment analysis, topic modelling, and identifying disfluencies in speech are prevalent.

Sentiment Analysis provides a framework for analyzing the valence (positive vs. negative) of a phrase.
Generally speaking, the sentiment of a phrase may not have a direct correlation with its rudeness, however sentiment analysis aims to give low scores to phrases which are perceived to be negative.
In~\cite{sentiment}, recursive neural networks are used to perform sentiment analysis and achieve more than 85\% accuracy in identifying the valence of a phrase.
There is likely to be much overlap between phrases of negative sentiment and offensive, rude speech, suggesting this approach can be applied to identify moments of rudeness.

Latent Dirichlet Allocation, a generative probabilistic model, can be used to identify inappropriate conversational topics~\cite{lda, sexdrugsviolence}.
These authors report their methods are able to achieve high recall (at least 0.94) for topics pertaining to sex and violence, however slightly weaker numbers for others.

Also, neural word embeddings are useful for identifying speech disfluencies (e.g., ``um", mid-sentence restarts, etc.)~\cite{blstm} and can achieve an F1 score greater than 0.85 using a bi-directional LSTM neural network on the Switchboard corpus of telephone conversation transcripts.
This approach is particularly interesting because conversational text is different from spoken language in that it does not contain speech disfluencies.
In theory, any algorithm which aims to successfully identify rude conversational behaviour must be robust against speech disfluencies.

Identifying moments of rude behaviour is directly complemented by identifying moments of politeness.
Recently, the authors in~\cite{wikipediapoliteness} annotate written text found in online communities and find strong correlations between the user's level of politeness towards others and his/her hierarchical rank in that community.
A linguistically informed classifier achieves at least 83\% accuracy on detecting rude demeanors in responses from forum users while incorporating the social rank as an additional variable.
This key finding reinforces the idea that individuals are more likely to be polite when they perceive themselves to be near the bottom of some structured hierarchy.
The politeness of a user's phrase is annotated using Amazon Mechanical Turk\footnote{\url{https://www.mturk.com}}.

\section{Methods}
We now define what it means for an individual to engage in rude conversational behaviour.
Fraser's argument against a universal grounding of politeness (and similarly, rudeness) suggests that being able to identify rudeness in speech, however we choose to define it, depends on the culture and context of speech.
For the purposes of this study, we consider a North American setting with English speakers.
Within this group, there may be mild variations in dialect (see Lakoff's example in section 1), however no significant differences in what is considered rude.

From now on, we use the terms \textit{user} and \textit{conversation partner} to refer to the individual whose conversational behaviour we wish to identify and the individual who the user is interacting with, respectively.
We define rude conversational behaviour to be all occurrences of \textbf{verbal insults}, \textbf{raised tones} (most commonly shouting), and \textbf{interruptions}.
Verbal insults comprise phrases whose semantic meanings are offensive; one common example is slandering.
Raised tones are intuitive and result from an increase in loudness during speech.
Interruptions occur when the user begins to speak while the conversation partner is still in the midst of formulating a sentence.

Note that there is a fourth category of rudeness which is also common in everyday settings: refusal of acknowledgement (i.e., ignoring), which we omit and do not attempt to identify.
This is because lengthy pauses between the time a question is posed and a response is provided may be misinterpreted as ignoring.
Also, rhetorical questions must be taken into consideration since they do not solicit responses.
In identifying refusal of acknowledgement, the identification of rhetorical questions is prerequisite, hence we only focus on the first three types of occurrences, which we shall refer to as the three classes of rude converational behaviour.

We assembled a dataset comprising audio clips of the three rudeness classes and the trivial class (non-rude).
In all, 67 audio clips of two-person conversations in which the user exhibits one of the three types of rudeness (or none) were collected from Hollywood films, popular TV shows and celebrity interviews.
Each audio clip is 10 seconds in length on average.
The data collection process was done manually; judgement of rudeness present in each example is that of the authors.
Some instances contain background noise such as people talking, music, etc. to create more practical scenario in which any real-time algorithm for detecting rude conversational behaviour should operate.
Laugh tracks are not present in the dataset as they are only present in certain TV shows.
In an attempt to diversify personalities and speech styles, we chose examples consisting of conversations between people of different ages, genders, occupations and ethnicities.
See table~\ref{dataset} for more details on the dataset.
An example transcript taken from \textit{The Sopranos}:

\begin{displayquote}
\textbf{U}: ``He's helping me to be a better catholic."\\
\textbf{CP}: ``Yeah, well we all got different needs."\\
\textbf{U}: ``What's different between you and me is you're going to hell when you die."
\end{displayquote}

Our experiments can be grouped into acoustic and semantic analyses.
These two approaches roughly correspond to the ``how" and ``what" of the user's speech: two major determinants of rude conversational behaviour.
For example, ``You're an asshole" can be said in a subtle tone so that acoustic signals may not reveal much about this verbal insult, whereas the semantic meaning makes all the difference.
In this case, the ``what" aspect is of interest.
Similarly, ``What's wrong with you?" is a question an elementary school teacher may ask a student who is feeling upset, however can be rude if the user shouts this question to the conversation partner in an indecent tone.
These two examples highlight the importance of acoustic and semantic analyses for identifying rude conversational behaviour.
Detecting interruptions relies mostly on semantic analysis and is later discussed in detail.

\begin{table}
\centering
\begin{tabular}{lcc}
\hline
Source & {\small Num. Examples} & {\small Num. Speakers} \\ \hline
The Departed & 8 & 5\\
Mean Girls & 13 & 5\\
Modern Family & 11 & 7\\
Tom Cruise Interview & 2 & 2\\
The Social Network & 10 & 7\\
Sons of Anarchy & 5 & 6\\
The Sopranos & 7 & 7\\
Suits & 1 & 2\\
{\small Vince McMahon Interview} & 1 & 2\\
Wolf of Wall Street & 9 & 7\\ \hline
\end{tabular}
\caption{Breakdown of out dataset by source, number of examples taken from that source, and the number of unique speakers in all audio clips from that source.}
\label{dataset}
\end{table}

\subsection{Acoustic Analysis}
Standard machine learning algorithms are used to classify different moments in the user's speech into one of four possible classes described above.
First, we train a model to identify different types of rudeness based on acoustic signals.
Similar to the approaches in~\cite{restroom, automatic}, we extract Mel-Frequency Cepstral Coefficients (MFCCs) from raw audio files at contiguous intervals separated by 10 milliseconds each.
This returns a 13-dimensional feature vector $(F_{1,i}, \ldots, F_{13,i})$ at each time frame $i$ which can then be used to train a Support Vector Machine (SVM) classifier.
However, we follow the same protocol in~\cite{restroom} by using the acceleration values of the MFCCs and omitting the first as this has proven to be superior towards distinguishing types of wave frequencies.
The features at time $i$ are $\mathbf{z}_i = (F''_{2,i}, \ldots, F''_{13,i})$.
We then train a SVM classifier using MFCC acceleration features extracted from raw audio files.
In addition, we make two additional modifications which may be beneficial: (1) only identifying instances of raised tones, and (2) using a two-tier classifier in which the first decides whether the user is being rude, and if so, the second the rudeness class.

In all cases, a smoothing function $h$ is used against the output of the SVM classifier at each time frame $i$, where $h(\mathbf{p}, i, w) = mode(\mathbf{p}_{i - \frac{w}{2}:i + \frac{w}{2}})$, where $\mathbf{p}$ is the discrete prediction vector of the SVM classifier.
At each timeframe $i$, the SVM classifier's output is a class, however in the context of classifying speech, it does not make sense for there to be high variance in class over a short window of length $10w$ ms.
For example, the user is highly unlikely to switch between engaging in rude conversational behaviour and then switching to an acceptable style many times over a few seconds.
Therefore, smoothing alleviates the problem of high variance output by taking a majority vote over all timeframes within a given window centered at the desired time frame.

Next, sound frequencies can also be used for interruption detection.
If the user is asking a question, his/her pitch is likely to be higher at the end of that question as compared with the beginning or middle of other sentences.
This is how voice intonation complements the semantics of how people speak.
We label each time frame when an individual is speaking as being part of either the beginning, middle, or end of a sentence or question.
A SVM classifier then learns to predict which part of a sentence a certain time frame belongs to based on MFCC values (similarly, we can perform clustering using $K$-means).
Once again, we use label smoothing to avoid high variance output over a short time window.

Lastly, in order to identify when the user is engaging in rude conversational behaviour, any autonomous system must know when he/she is speaking and not confuse the conversation partner to be the user.
We use a feed-forward neural network with the architecture presented in~\cite{neuralnetworkdiarization} to perform speaker \textit{diarization}: the process of partitioning audio based on the speaker at that time.
The network takes two windows of MFCC acceleration features $\mathbf{W}_t$, $\mathbf{W}_{t'}$ as input, where $\mathbf{W}_t$ $=$ $\{\mathbf{z}_t$, $\mathbf{z}_{t+1}$, $\ldots$, $\mathbf{z}_{t+M-1}\}$ and $\mathbf{W}_{t'}$ $=$ $\{\mathbf{z}_{t'}$, $\mathbf{z}_{t'+1}$, $\ldots$, $\mathbf{z}_{t'+M-1}\}$.
$M$ is the length of the window of features.
The network's objective is to determine whether the two sets of MFCC acceleration features $\mathbf{W}_t, \mathbf{W}_{t'}$ are produced by the same speaker (i.e., if the speaker at time $t$ is same as the speaker at time $t'$).
Note that a window of MFCC acceleration features is used by the network to discriminate, since $\mathbf{z}_t$ and $\mathbf{z}_{t'}$ are sounds at unique time frames and likely insufficient to determine if the same speaker is produces both sounds.
Speaker diarization can also be useful for detecting interruptions, assuming methods in semantic analysis can determine when the conversation partner started saying a phrase which is incomplete.

In our study, speaker diarization is used in lieu of speaker identification to allow the model to generalize.
In all, our dataset has 50 distinct speakers.
We would like to distinguish between speaker in scenarios when the model is not familiar with at least one of the speakers.
For instance, the model does not learn to identify a conversation partner, however must do exactly this at some subsequent time.
Speaker diarization is more appropriate in theory since the model learns to distinguish between different speakers based on their perceived acoustic differences.

\subsection{Semantic Analysis}
Our first approach to analyzing content in speech is with a Naive Bayes (NB) classifier~\cite{naivebayes}.
This approach is ideal for distinguishing between rude and acceptable phrases based on semantic content.

Second, we use the Stanford Sentiment Analysis tool~\cite{sentiment} to assign positive or negative valences to content using a pre-trained deep recursive neural network.
This method constructs a tree of words such that flattening the tree would result in a linear sequence of words which is the same as the original phrase.
The words are then fed into a recursive neural network from the leaves upwards which assigns a sentiment score.
The final score is then used to categorize the phrase as \textit{very negative}, \textit{negative}, \textit{neutral}, \textit{positive} or \textit{very positive}.

\section{Empirical Results}

\subsection{Classification using SVM}
In total, our dataset of audio clips translates to roughly 65,000 time frames of MFCC acceleration feature vectors.
This is enough to train a 4-way SVM classifier.
We train models with different kernel types (Gaussian vs. polynomial) and parameters.
In all experiments, we scale the values of each $\mathbf{z}_i$ to be between 0 and 1 as we found this improves classification accuracy.
The size of the smoothing window is $w=30$ (i.e., 0.3 seconds).
LIBSVM~\cite{libsvm} is used to implement all models.

Any arbitrary classifier can achieve an accuracy of approximately 70\% by trivially predicting that the user is not being rude at each moment during a one-on-one interaction.
This is because in most conversations where the user engages in some kind of rude conversational behaviour, roughly less than one-third of time frames consist of rudeness.
As shown in table~\ref{acoustic-results}, our best model achieves just under 78.9\% accuracy, a slight improvement over the baseline.
However, a more interesting and relevant question for our task is how well the same model performs on detecting instances of rudeness.
Surprisingly, each model's ability to correctly identify instances of rudeness is quite poor compared to its overall accuracy.
The top-performing model on rudeness classes achieves just 26.4\% accuracy.

Next, we try detecting only raised tones (i.e., we are only interested in identifying one type of rudeness class).
We posit that verbal insults and interruptions are perhaps more difficult to capture through MFCCs than raised tones, which are more likely to be perceptible to a classifier due to an increase in pitch and volume.
The best model in this experiment is able to correctly classify 41.4\% of raised tone instances (see figure~\ref{acoustic-results-shouting}).
Some models (i.e., Gaussian kernel with $\gamma=1.0$ and linear kernel) never believes the user is speaking with a raised tone, suggesting these models fail to learn accurate representations of MFCC acceleration features for this task.

Finally, we try a two-tier classifier to identify instances of rude conversational behaviour.
The second-tier 3-way classifier distinguishes when the user is speaking with a raised tone better than any other type of rudeness.
The first-tier classifier often performs poorly, leaving the second-tier classifier to choose between one of three classes of rudeness where many MFCC acceleration features are false positives.

\begin{table}
\centering
\noindent\begin{tabular}{l|c|c|c|c|}
 \multicolumn{1}{c}{} & \multicolumn{1}{c}{None} & \multicolumn{1}{c}{Insult} & \multicolumn{1}{c}{R. tone} & \multicolumn{1}{c}{Interr.} \\ \hhline{~*4{|-}|}
 None & 0.77 & 0.07 & 0.02 & 0.01 \\ \hhline{~*4{|-}|}
 Insult & 0.02 & 0.01 & 0.01 & 0 \\ \hhline{~*4{|-}|} 
 R. tone & 0.04 & 0.01 & 0.02 & 0 \\ \hhline{~*4{|-}|}
 Interr. & 0.01 & 0 & 0.01 & 0 \\ \hhline{~*4{|-}|}
\end{tabular}\par\bigskip
\caption{The confusion matrix of a single-tier classifier using a Gaussian kernel with $\gamma=0.5$.}
\label{confusion}
\end{table}

{\footnotesize
\begin{table}
\centering
\begin{tabular}{l*{3}{c}}
\hline
 \multicolumn{1}{c}{} & \multicolumn{2}{c}{Accuracy (\%)}\\ 
 \multicolumn{1}{c}{SVM kernel} & \multicolumn{1}{c}{All Classes} & \multicolumn{1}{c}{Rude Classes} \\ \hline
 \textbf{Gaussian, $\gamma=0.05$} & 68.0 & \textbf{26.4} \\
 \textbf{Gaussian, $\gamma=0.5$} & 75.6 & 2.7 \\
 \textbf{Gaussian, $\gamma=1.0$} & 84.0 & 0 \\ 
 \textbf{Polynomial, degree 1} & 73.6 & 0 \\
 \textbf{Polynomial, degree 3} & \textbf{78.9} & 10.6 \\
 \textbf{Polynomial, degree 6} & 50.6 & 22.9 \\ \hline
\end{tabular}
\caption{Accuracies from training different models to classify each time frame into one of three classes of rudeness or none.}
\label{acoustic-results}
\end{table}
}

\begin{table}
\centering
\begin{tabular}{l*{2}{c}}
\hline
 \multicolumn{1}{c}{} & \multicolumn{2}{c}{Accuracy (\%)}\\ 
 \multicolumn{1}{c}{SVM kernel} & \multicolumn{1}{c}{Regular} & \multicolumn{1}{c}{Smoothed} \\ \hline
 \textbf{Gaussian, $\gamma=0.05$} & 68.0 & 26.4 \\
 \hspace*{6mm}Raised Tones & 35.0 & 13.8\\
 \hspace*{6mm}Other & 95.1 & 99.5\\ \hline
 \textbf{Gaussian, $\gamma=0.5$} & 87.5 & 87.8 \\
 \hspace*{6mm}Raised Tones & 0.3 & 0\\
 \hspace*{6mm}Other & 99.6 & 100\\ \hline
 \textbf{Gaussian, $\gamma=1.0$} & 88.1 & 88.1 \\ 
 \hspace*{6mm}Raised Tones & 0 & 0\\
 \hspace*{6mm}Other & 100 & 100\\ \hline
 \textbf{Polynomial, degree 1} & 96.0 & 96.0 \\
 \hspace*{6mm}Raised Tones & 0 & 0\\
 \hspace*{6mm}Other & 100 & 100\\ \hline
 \textbf{Polynomial, degree 3} & 90.7 & 90.8 \\
 \hspace*{6mm}Raised Tones & 11.2 & 0\\
 \hspace*{6mm}Other & 98.8 & 100\\ \hline
 \textbf{Polynomial, degree 6} & 72.9 & 86.1 \\
 \hspace*{6mm}Raised Tones & \textbf{41.4} & \textbf{20.2}\\
 \hspace*{6mm}Other & 76.9 & 94.6\\ \hline
\end{tabular}
\caption{Results from performing binary classification to detect }
\label{acoustic-results-shouting}
\end{table}

\subsection{Detecting Insults with Naive Bayes}
Just as we used a SVM classifier on MFCC acceleration features to identify raised tones, we use a NB classifier for binary classification once again.
This time, however, we are interested in detecting verbal insults since a NB classifier is designed for extracting semantic content.
Our experiments show the classifier adapts poorly to the task and is unable to break a threshold of 70\% accuracy.
Potential causes for this are discussed in subsequent sections.

\subsection{Speaker Diarization}
In our experiments, we use $M=10$ (i.e., 0.1 second windows) while $t$ and $t'$ are separated by one second.
We pair windows of MFCC acceleration features $\mathbf{W}_t, \mathbf{W}_{t+1}$ and train a neural network to identify when the speaker has changed using the architecture described in the previous section.

Similar to the case of classification using SVMs, speakers are less likely to change frequently over a short time interval.
The output of a neural network may have high variance over a short-time period, and applying a smoothing function $h$ may not entirely alleviate this problem, as evident in our results.
Modelling the speaker through a hidden state as in a Hidden Markov Model would perhaps better conform with the nature of speaker diarization.

Moreover, measuring accuracy in the binary classification sense (same speaker versus new speaker) is inherently different.
A model performing speaker diarization can achieve accuracy greater than 90\%, however the number of speaker changes can deviate greatly the actual number of speaker changes.
As such, percentage accuracy is an incorrect measure of the strength of a speaker diarization classifier and this reinforces the need for different type of model -- such as a finite state model.

\subsection{Sentence Segmentation}
\begin{figure}
\centering
\begin{minipage}{0.45\textwidth}
\centering
\includegraphics[width=2.35in]{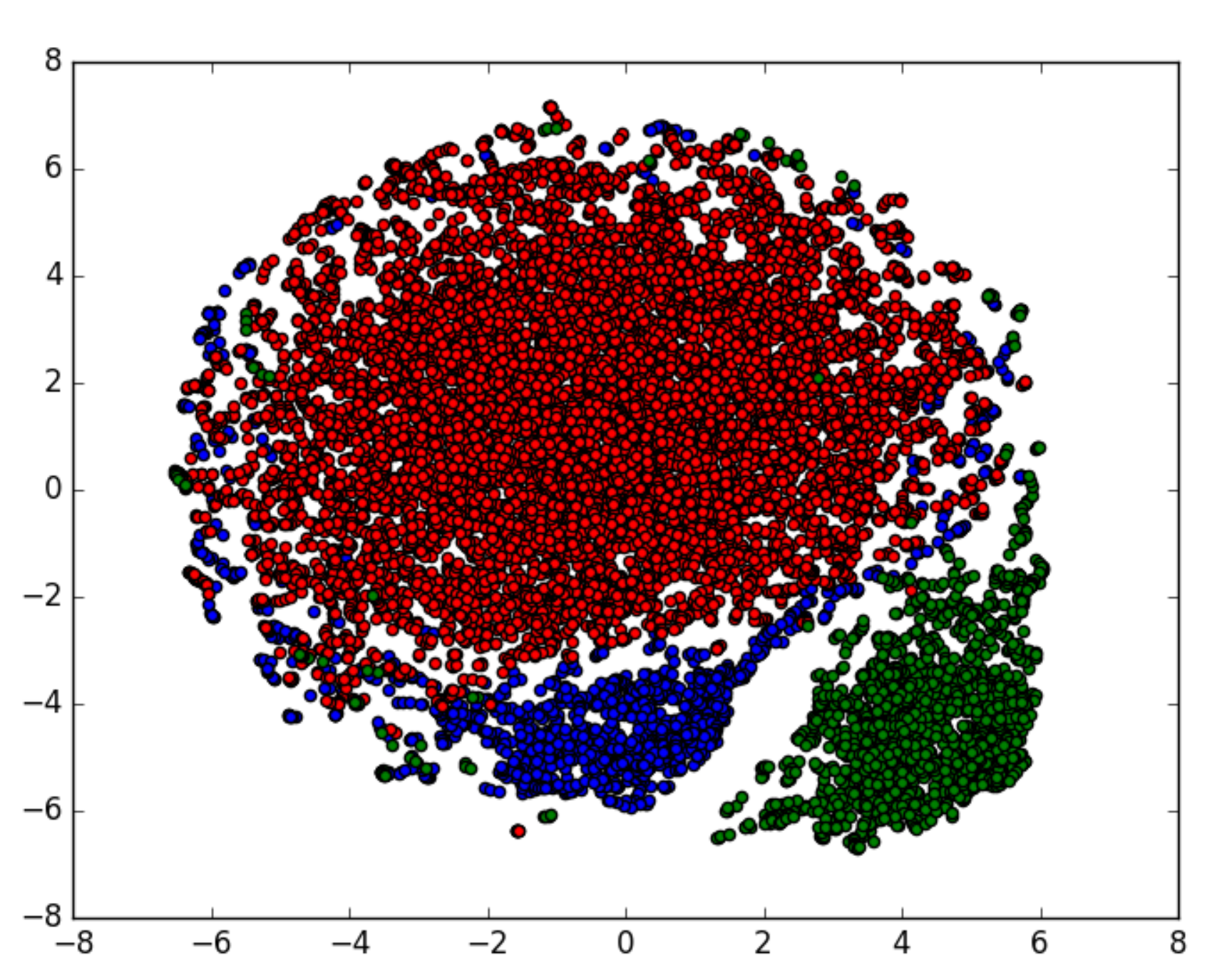}

(a) t-SNE
\end{minipage}
\begin{minipage}{0.45\textwidth}
\centering
\includegraphics[width=2.35in]{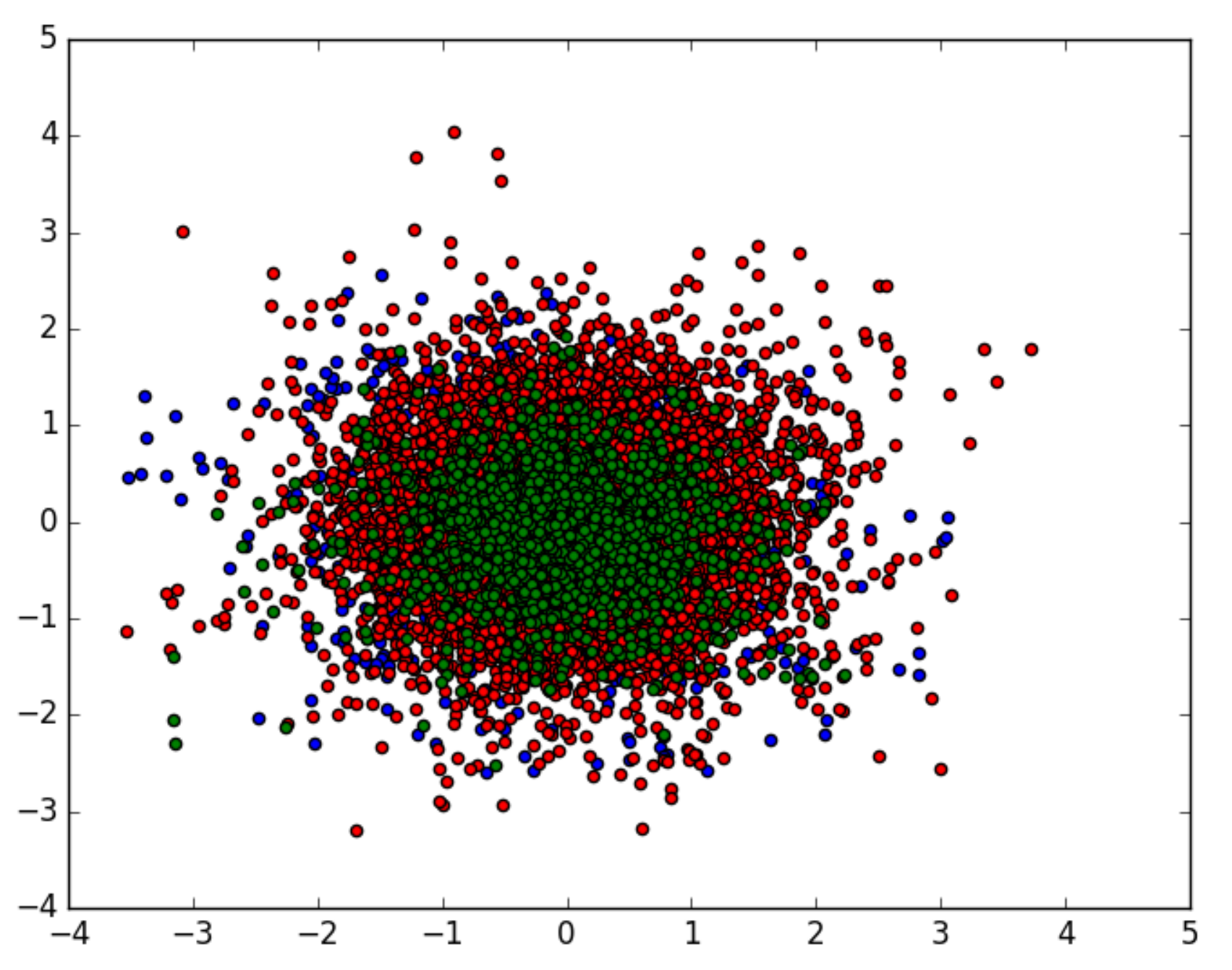}

(b) PCA
\end{minipage}
\caption{The dimensionality reduction using t-SNE suggests the beginning, middle and end of phrases are distinguishable with respect to their MFCC acceleration values; the results from PCA contradict this view.}
\label{tsne-vs-pca}
\end{figure}

We define the beginning and end of a phrase\footnote{Here we use the term \textit{phrase} synonymously with sentence and question.} to be the first and last words in that phrase, respectively; the rest is the middle.
Note that the conversation partner's phrases which are interrupted mid-sentence by the user do not have an end -- just a beginning and middle since the intended phrase is incomplete.
To determine whether different parts of sentence are significantly different with respect to MFCC acceleration values, we visualize the features via dimensionality reduction.
Figure~\ref{tsne-vs-pca} compares results using both t-SNE~\cite{tsne} and Principal Component Analysis (PCA).
The t-SNE results suggests different parts of sentence should be distinguishable to kernel methods, however the PCA results contradict this view.

We train a SVM classifier and $K$-means clustering algorithm (where $K=3$) to distinguish between the beginning, middle and end of each phrase.
The results in table~\ref{segmentation-results} illustrate how, despite the promising results from the t-SNE, no SVM classifier is able to perform well on all three classes simultaneously.
A SVM classifier with a Gaussian kernel and $\gamma=0.5$ achieves almost 80\% accuracy on correctly identifying the end of the phrase, but is only able to identify the middle correctly about half the time.
As mentioned in the previous section, the motivation for correctly identifying the end is to determine if a phrase is complete; if not, there is a chance that an interruption occurred.
$K$-means also performs poorly, even when the number of instances which can be classified as a beginning, middle or end are restricted to satisfy proportions from the dataset.
We restrict portions since the distribution between parts of a phrase is clearly not equal.

\begin{table}
\centering
\begin{tabular}{l*{2}{c}}
\hline
 \multicolumn{1}{c}{} & \multicolumn{2}{c}{Accuracy (\%)}\\ 
 \multicolumn{1}{c}{SVM kernel} & \multicolumn{1}{c}{Regular} & \multicolumn{1}{c}{Smoothed, $w=15$} \\ \hline
 \textbf{Gaussian, $\gamma=0.05$} & 47.6 & 50.2 \\
 \hspace*{6mm}Beginning & 5.0 & 0\\
 \hspace*{6mm}Middle & 46.8 & 51.0\\
 \hspace*{6mm}End & \textbf{78.9} & \textbf{74.3}\\ \hline
 \textbf{Gaussian, $\gamma=0.5$} & 58.3 & 71.2 \\
 \hspace*{6mm}Beginning & \textbf{10.9} & 4.0\\
 \hspace*{6mm}Middle & 63.6 & 80.2\\
 \hspace*{6mm}End & 47.4 & 45.0\\ \hline
 \textbf{Gaussian, $\gamma=1.0$} & 64.3 & 77.5 \\ 
 \hspace*{6mm}Beginning & 6.9 & \textbf{5.0}\\
 \hspace*{6mm}Middle & \textbf{73.1} & \textbf{90.2}\\
 \hspace*{6mm}End & 33.9 & 26.9\\ \hline
\end{tabular}
\caption{Classification results for identifying the beginning, middle, and end of a phrase. SVM classifiers with polynomial kernels perform remarkably worse and are omitted.}
\label{segmentation-results}
\end{table}

\subsection{Sentiment Analysis}
Sentiment analysis is only slightly more effective when used to score sentiments of speech with offensive or inappropriate semantic meaning as opposed to speech with raised tones or interruptions.
We passed a sample of conversation transcripts from our collected dataset to the Stanford Sentiment Analysis tool which then assigned a single score to the entire transcript (recall a transcript is only worth 10 seconds of conversation, on average).
The results in table~\ref{sentiment-analysis-results} demonstrate how sentiment analysis assigns more \textit{negative} scores to conversations with offensive or inappropriate content, however the margin is quite low.
Nonetheless, less than half of the transcripts with offensive or inappropriate content are assigned a \textit{negative} score.
Content in the \textit{non-insults} category comprises phrases which may not be directly offensive from a semantic perspective, so the difference in results is not strong enough to conclude sentiment analysis is effective towards insults.

\begin{table}
\centering
\begin{tabular}{lccc}
\hline
Class & \% negative & \% neutral & \% positive \\ \hline
Verbal Insults & 46.9 & 37.5 & 15.6\\
Other & 40.0 & 43.3 & 16.7\\
\hline\end{tabular}
\caption{Classification using Sentiment Analysis on a subset of the collected data.}
\label{sentiment-analysis-results}
\end{table}

\section{Discussion}
We introduced a dataset consisting of short audio clips in which the user is being rude to their conversation partner.
In an effort to expand our collection, we encourage researchers to contribute to our dataset\footnote{See \url{http://www.cs.toronto.edu/~grewal/rudeness.html} for dataset contributions.} in order to throughly train classification models.

In this study, we looked at acoustic and semantic analyses for the purpose of identifying rude conversational behaviour.
We showed a SVM classifier is better at determining when a user is speaking with a raised tone than deciding on other classes of rudeness.
This is because shouting, for example, can be distinguished easily from regular speech based solely on acoustics. On the other hand, verbal insults and interruptions may not be acoustically discernable from regular speech.
This analysis corresponds to the ``how" aspect of what the user says.
However, the classifier is only able to correctly identify moments of raised tones on fewer than half those instances -- why is performance so bad?
An important consideration is whether we are using relevant data to perform acoustic analysis.
Raised tones may be a function of facial expression, prosody, and MFCCs together.
By looking only at MFCCs, we may be ignoring important information about the user's behaviour, hence poor classification accuracy
In short, our study only looked at MFCCs and this may be a limiting factor.

The NB method is a statistical method, and like most others, relies heavily on sufficient data.
The presence of only a few-hundred sentences in our dataset may pose a problem.
An evident interpretation of why NB performs poorly is the following: $V$ is the vocabulary of words in our dataset and $\mathbf{x}$ a binary vector of size $|V|$ which represents a phrase, where the $i^{\text{th}}$ component $\mathbf{x}_i = 1$ if and only if word $j$ appears in the phrase and $0$ otherwise.
The posterior probability of a given phrase $\mathbf{x}$ belonging to class $C$ is proportional to $\Pi_i p(\mathbf{x}_i | C)$ (where $C$ is binary: \textit{rude} or \textit{not rude}).
If some words do not occur in one of the classes, the product of likelihoods is zero.
Consider, for instance, the word ``stupid" which occurs five times in a rude phrase and never in a non-rude phrase in the collected dataset.
Then, $p(\mathbf{x}_k | C=\text{Not Rude}) = 0$ and so $p(C=\text{Not Rude} | \mathbf{x}) = 0$ (where "stupid" corresponds to the $k^{\text{th}}$ word in the vocabulary) by product of likelihoods.
In reality, however, we know this is clearly not the case, as the user may demonstrate positive conversational behaviour despite using the word ``stupid".
Our dataset is thus not diverse enough to apply a NB classifier.
In particular, we need a dataset such that all words in vocabulary $V$ should appear in instances of both rude and non-rude phrases as it would ensure a more accurate measure of semantic rudeness and verbal insults.

Measuring the accuracy of speaker diarization in the binary classification sense (same speaker versus new speaker) is a poor choice of metric to determine the strength of a model.
For instance, a model performing speaker diarization can achieve accuracy greater than 95\% accuracy, however the number of speaker changes  can  deviate  greatly  the  actual  number  of  speaker changes.
As such, percentage accuracy is an incorrect measure  of  the  strength  of  a  speaker  diarization  classifier  and this reinforces the need for different type of model.
Instead, a finite state model used for identifying when the user is speaking will remember how long the speaker at any moment has been speaking for (typically follows a distribution) and this information can be combined with sentence segmentation to determine when the speaker is likely near the end of a phrase.

Our experimental results demonstrate that identifying interruptions is far from solved.
The motivation for performing sentence segmentation is to combine this technique with speaker diarization/identification and ultimately determine when the user abruptly cuts off the conversation partner.
Until we are able perform both subtasks with moderate accuracy, we will not be able to perform interruption detection.

\section{Conclusions and Future Work}
Identifying when the user is behaving inappropriately during a conversation is a difficult task.
In this work, we demonstrate that tools from the signal processing and natural language domains can be applied and used in tandem to detect rude conversational behaviour.
Also, we provide explanations for why our methods perform poorly.
We recommend the following to researchers who aim to tackle this very task:

\begin{itemize}
\item Stronger statistical methods such as a maximum entropy classifier to discern rude semantic content.
\item Using a finite state model (e.g., Hidden Markov Model) to identify whether the user or conversation partner is speaking at any given time. A ``same vs. different" speaker diarization approach is inherently more challenging.
\item Combining knowledge of who is speaking (user or conversation partner) with occurrences of incomplete phrases to accurately identify interruptions.

\end{itemize}
In the future, a promising direction motivated by the social applications discussed in section 1 is to develop a ubiquitous computing application which can intervene to inform the user of any undesirable or inappropriate actions on his/her part.

\bibliographystyle{abbrv}
\bibliography{main}

\end{document}